\documentclass[%
 reprint,
 amsmath,amssymb,
 aps,
]{revtex4}

\usepackage{graphicx}% Include figure files
\usepackage{dcolumn}% Align table columns on decimal point
\usepackage{bm}% bold math
\begin{document}

\preprint{APS/123-QED}

\title{Spontaneous Scalarization of Charged Gauss-Bonnet Black Holes: Analytic Treatment}% Force line breaks with \\

\author{Shun Jiang}

\email{shunjiang@mail.bnu.edu.cn}

\affiliation{%
	Department of Physics, Beijing Normal University, Beijing, 100875, China\\}%Lines break automatically or can be forced with \\

\date{\today}% It is always \today, today,

\begin{abstract}
Recently, by considering nontrivial couplings between scalar fields and the Gauss-Bonnet invariant, the Schwarzschild black hole may allow regular scalar hairy configurations. The numerical studies of this spontaneous scalarization phenomenon show if the coupling parameter $\bar{\eta}$ belongs to discrete sets $\bar{\eta}\in\{\bar{\eta}^-_n,\bar{\eta}^+_n\}^{n=\infty}_{n=0}$, the black hole can support regular scalar hairy configurations. Interestingly, Hod finds the coupling parameter $\bar{\eta}^+_n$ which correspond to the black hole linearized scalar field configurations has an asymptotic universal behavior  $\Delta_n\equiv\sqrt{\bar{\eta}_{n+1}}-\sqrt{\bar{\eta}_{n}} \simeq2.72$. He provides a remarkably compact analytical explanation for the numerically observed universal behavior. Motivated by this interesting phenomenon, in this paper, we study the coupling parameter behavior in RN black hole with a nontrivial coupling between scalar fields and the Gauss-Bonnet invariant. Different from Schwarzschild case where the coupling parameter can only take positive values, in this case, the coupling parameter can take positive or negativie values. Therefore, it is interesting to investigate whether the coupling parameter has a similar asymptotic behavior in this situation. By examining numerical data, we find there is a similar asymptotic behavior for both positive and negative parameters. We also compare the analytical results with the numerical data. We find analytical results agree well with the the numerical data. 
\end{abstract}

\maketitle

%\tableofcontents

\section{introduction}
The direct detections of gravitational waves\cite{A1,A2,A3} provide the evidence of the existence of black hole. With the increasing of these detections, one may examine whether there are some characteristics which is different from general relativity. For black hole case, there may have some fundamental scalar fields interating with black holes. However, no-hair theorems dictate that asymptotically flat black holes in general relativity can't support spatially regular static scalar field configurations \cite{Bekenstein:1971hc,Mayo:1996mv,Sotiriou:2011dz,Sotiriou:2015pka,Herdeiro:2015waa,Hod:2019sjj,Hod:2017ssh}. This become an obstacle to consider new fundamental scalar field.

Recently, the no-hair theorem is violated by considering a non-minimally coupling of scalar field and Gauss-Bonnet invariant \cite{Sotiriou:2013qea,Doneva:2017bvd,Silva:2017uqg,Antoniou:2017acq}. In particular, they show Gauss-Bonnet term may lead the Schwarzschild black hole become unstable and the corresponding black hole with hair is more stable. Therefore, the Schwarzschild black hole will evolve into the corresponding hair black hole. This physical phenomenon is called spontaneous scalarization which was first found in neutron stars \cite{Damour:1993hw}. Furthermore, spnontanous scalarization was also found in Kerr black hole \cite{Cunha:2019dwb,Herdeiro:2020wei}. Since then, various black hole with non-minimally coupling of scalar field and Gauss-Bonnet invariant has been studied  \cite{Brihaye:2018bgc,Macedo:2019sem,Brihaye:2019gla,Herdeiro:2019yjy,Brihaye:2019dck,Brihaye:2019kvj}. Interestingly, spontaneous scalarization can also be be induced by Maxwell fields when one consider non-minimal couplings between scalar fields and Maxwell fields \cite{Fernandes:2019rez,Herdeiro:2018wub}. At present, spontaneous scalarization phenomenons was also studied in various gravity theory \cite{Brihaye:2018acx,Doneva:2018rou,Motohashi:2018mql,Minamitsuji:2018xde,Zou:2019bpt,Blazquez-Salcedo:2018jnn,Myung:2018vug,Myung:2019oua,Peng:2019cmm,Zou:2019ays,Herdeiro:2020htm}. In addition, analytic treatment of spontaneous scalarization were studied in \cite{Hod:2019pmb,Hod:2019vut,Hod:2020ljo,Peng:2020znl,Hod:2020jjy}. 

When studying spontaneous scalarization phenomenons, it is important to find the ranges of coupling parameter which allow hair black hole solutions. In \cite{Doneva:2017bvd,Silva:2017uqg}, they consider non-minimally coupling of scalar field and Gauss-Bonnet invariant in Schwarzschild black and the numerical computations show the spontaneous scalarization is characterized by discrete sets $\eta\in\{\eta^-_n,\eta^+_n\}^{n=\infty}_{n=0}$. In particular, the discrete eigenvalues $\{\eta^+_n\}^{n=\infty}_{n=0}$ correspond to composed black-hole-linearized-scalar-field cloudy configurations. In other words, this correspond to test field limit where the background is fixed. Interestingly, in \cite{Hod:2019pmb}, by examing the numerical data that are available in the physics literature, they find the coupling parameter $\eta^+_n$ has an asymptotic universal behavior $\sqrt{\eta^+_{n+1}}-\sqrt{\eta^+_n}\simeq2.72$. Motivated by this interesting observation, they use an analytic method to study the $\eta^+_n$'s asymptotic behavior. Surprisingly, the numerical data are described extremely well by the analytic method(the deviations are smaller than $1\%$ in the large $n$ limit). Motivated by it, it is interesting to investigate whether there is a similar asymptotic universal behavior in other black holes. Therefore, in this paper, we consider the RN black hole with a non-minimally coupling of scalar field and Gauss-Bonnet invariant. This model has been numerical studied in \cite{Brihaye:2019kvj}. Different from Schwarzschild black hole where the coupling parameter can only take positive value, in this case, the coupling parameter can take positive or negative value. Therefore, we will use the analytic method to study the coupling parameter's asymptotic behavior for both positive and negative cases. To check the validity of this analytic method, we will also compare analytic results with the numerical data.  

The rest of this paper is as follows. In Sec.~\ref{seciton2}, we introduce the model with a scalar field coupled to Gauss-Bonnet invariant. In Sec.~\ref{section3}, we use the analytic method to study the coupling parameter's asymptotic behavior. In Sec.~\ref{ssss}, we compare analytic results with the numerical data. In Sec.~\ref{seciton5}, we give a brief conclusion.
\section{DESCRIPTION OF THE MODEL}\label{seciton2}
In this paper, we consider the model contains a non-minimal coupling between a real scalar field and the Gauss-Bonnet term. The model reads \cite{Brihaye:2019kvj}
\begin{eqnarray}
	S=\int d^4x\sqrt{-g}[\frac{R}{2}+\gamma\phi^2\mathcal{R}_{GB}^{2}-\partial_a\phi\partial^a\phi-\frac{1}{4}F^{\mu\nu}F_{\mu\nu}]
\end{eqnarray}
Where $\gamma$ is the coupling parameter and the Gauss-Bonnet term $\mathcal{R}_{GB}^{2}$ is given by
\begin{eqnarray}
	\mathcal{R}_{GB}^{2}=R_{abcd}R^{abcd}-4R_{ab}R^{ab}+R^2
\end{eqnarray}
We use the units in \cite{Brihaye:2019kvj} where $8\pi G=1$. Using the conventions in \cite{Brihaye:2019kvj}, the RN solution reads
\begin{eqnarray}
	ds^2=-Ndt^2+\frac{1}{N}dr^2+r^2(d\theta^2+\sin^2\theta d\varphi^2)
\end{eqnarray}
\begin{eqnarray}
	N(r)=1-\frac{2M}{r}+\frac{Q^2}{2r^2}
\end{eqnarray}
The event horizon is located at $r_h=M+\sqrt{M^2-Q^2/2}$ with the extremal limit at $r_h=M={Q}/\sqrt{2}$. Variation with respect to the scalar field $\phi$ leads to the following equations
\begin{eqnarray}
\partial_a\partial^a\phi+\gamma\phi\mathcal{R}_{GB}^{2}=0 \label{scalar eq}
\end{eqnarray}
Using the field decomposition
\begin{eqnarray}
	\phi(r,\theta,\varphi)=\sum\frac{\psi_{lm}}{r}Y_{lm}(\theta,\varphi)
\end{eqnarray}
Eqs.~(\ref{scalar eq}) can be written as
\begin{eqnarray}
	N\frac{d^2\psi}{dr^2}+N'\frac{d\psi}{dr}=(\frac{l(l+1)}{r^2}+\frac{N'}{r}-\gamma\mathcal{R}_{GB}^{2})\psi \label{scalar eqq}
\end{eqnarray}
We use the tortoise coordinate $y$ defined by
\begin{eqnarray}
	\frac{dr}{dy}=N(r)
\end{eqnarray}
We obtain a Schr\"{o}dinger-like differential equation
\begin{eqnarray}
	\frac{d^2\psi}{dy^2}-V\psi=0 \label{SSS}
\end{eqnarray}
The effective radial potential in the Schr\"{o}dinger-like differential equation is given by
\begin{eqnarray}
	V(r)=N(\frac{l(l+1)}{r^2}+\frac{N'}{r}-\gamma\mathcal{R}_{GB}^{2})
\end{eqnarray}
In \cite{Brihaye:2019kvj}, they solve this problems with including backreaction of scalar field and the solutions of coupling parameter which allow hairy black hole are many discrete intervals. For $\gamma>0$, the hairy black hole is characterized by a discrete set $\gamma\in{[\gamma^+_{cr(n)},\gamma^+_{0(n)}]}^{n=\infty}_{n=0}$. For $\gamma<0$, the hairy black hole is characterized by a discrete set $\gamma\in{[\gamma^-_{0(n)},\gamma^-_{cr(n)}]}^{n=\infty}_{n=0}$. In particular, the endpoints $\gamma_{0(n)}$ correspond to the static bound-state scalar field configurations and it can be determined by the differential equation (\ref{SSS}) with the boundary condition $\psi(r_h<\infty)$ and $\psi(\infty)\rightarrow0$. In Schwarzschild black hole, they show the coupling parameter exists an asymptotic universal behavior. Therefore, in next section, we will use the analytic method to explore whether the discrete resonant spectrum $\{\gamma_{0(n)}\}^{n=\infty}_{n=0}$ has a similar asymptotic universal behavior.

\section{analytic method}\label{section3}
In this section, we use the analytic method to solve the Schr\"{o}dinger-like differential equation (\ref{SSS}). The Schr\"{o}dinger-like differential equation (\ref{SSS}) can be studied by a standard WKB analysis. For bound-state field configurations, there is a well-known quantization condition
\begin{eqnarray}
	\int_{y_-}^{y_+}\sqrt{-V(y,\gamma)}dy=(n-\delta)\pi; \qquad  n=1,2,3,...., \label{V}
\end{eqnarray}
where $\{y_-,y_+\}$ are the turning point of $V(y)$. $\delta$ is phase depends on the number of turing points.
Using the relation between tortoise coordinate $y$ and radial coordinate $r$, Eqs.~(\ref{V}) can be written as
\begin{eqnarray}
	\int_{y_-}^{y_+}\frac{\sqrt{-V(y,\gamma)}}{N(r)}dr=(n-\delta)\pi; \qquad  n=1,2,3,...., \label{VP}
\end{eqnarray}
It is easy to see horizon radius $r_h$ leads $V(r_h)=0$. Therefore, the horizon radius $r_h$ is a turning point. The other turning points is determined by
\begin{eqnarray}
\frac{l(l+1)}{r^2}+\frac{N'}{r}-\gamma\mathcal{R}_{GB}^{2}=0 \label{GB}
\end{eqnarray}
In the eikonal large $\gamma$ regime, the Eqs.~(\ref{GB}) can be approximated by
\begin{eqnarray}
\mathcal{R}_{GB}^{2}=0
\end{eqnarray}
Using this approximation, Eqs.~(\ref{VP}) can be written as
\begin{eqnarray}
	\int_{r_-}^{r_+}\sqrt{\frac{{\gamma \mathcal{R}_{GB}^{2})}}{N(r)}}dr=(n-\delta)\pi;\qquad  n=1,2,3,...., \label{plusa}
\end{eqnarray}
In order to find the turning points, we need to study the sign of $\mathcal{R}_{GB}^{2}$ at the interval $[r_h,\infty]$. This has been done in \cite{Brihaye:2019kvj}. We will briefly review the conclusion. The expression of $\mathcal{R}_{GB}^{2}$ reads \cite{Brihaye:2019kvj}
\begin{eqnarray}
	\mathcal{R}_{GB}^{2}=\frac{12}{r^6}+\frac{12(r-2)Q^2}{r^7}+\frac{(3r^2-12r+10)Q^4}{r^8} \label{T}
\end{eqnarray}
where they set $r_h=1$ and use $Q$ to replace $M$ in expression of $\mathcal{R}_{GB}^{2}$.

For $Q<\tilde{Q}=\sqrt{2(3-\sqrt{6})}\approx1.05$. The $\mathcal{R}_{GB}^{2}$ is postive at at the interval $[r_h,\infty]$. Therefore, the turning point are $r_-=r_h$ and $r_+=+\infty$.  From Eqs.~(\ref{plusa}), we find the $\gamma$ only can take postive value in this situation. Eqs.~(\ref{plusa}) becomes
\begin{eqnarray}
\sqrt{\gamma_{0(n)}}=(n-\frac{1}{4})\pi/\int_{r_h}^{+\infty}\sqrt{\frac{{\mathcal{R}_{GB}^{2})}}{N(r)}}dr;\qquad  n=1,2,3,....,\label{1}
\end{eqnarray}

For $\sqrt{2}>Q>\tilde{Q}$. There is a turning point $r_0$ where $\mathcal{R}_{GB}^{2}(r_0)=0$ and $r_h<r_0<\infty$. Therefore, the $\mathcal{R}_{GB}^{2}(r)$ can be divided into two parts: $\mathcal{R}_{GB}^{2}(r)<0$ at $(r_h,r_0)$ and $\mathcal{R}_{GB}^{2}(r)>0$ at $(r_0,+\infty)$. Therefore, from Eqs.~(\ref{plusa}), we see the $\gamma$ can take postive or negative value. This is different from Schwarzschild case where $\gamma$ can only take positive value.

For positive $\gamma$, we take $r_-=r_0$, $r_+=+\infty$ and Eqs.~(\ref{plusa}) becomes
\begin{eqnarray}
	\sqrt{\gamma_{0(n)}}=(n-\frac{1}{4})\pi/\int_{r_0}^{+\infty}\sqrt{\frac{{\mathcal{R}_{GB}^{2})}}{N(r)}}dr;\qquad  n=1,2,3,....,\label{2}
\end{eqnarray}
For negative value $\gamma$, we take $r_-=r_h$,$r_+=r_0$ and Eqs.~(\ref{plusa}) becomes
\begin{eqnarray}
	\sqrt{-\gamma_{0(n)}}=(n-\frac{1}{2})\pi/\int_{r_h}^{r_0}\sqrt{\frac{{\mathcal{R}_{GB}^{2})}}{N(r)}}dr;\qquad  n=1,2,3,....,\label{3}
\end{eqnarray}
Therefore, we get three expressions of $\gamma_{0(n)}$ in different situations. Using their discrete spectrum, we can find quantity relations of their behavior

For $Q<\tilde{Q}=\sqrt{2(3-\sqrt{6})}\approx1.05$, from Eqs.~(\ref{1}), we find
\begin{eqnarray}
	\sqrt{\gamma_{0(n+1)}}-\sqrt{\gamma_{0(n)}}=\pi/\int_{r_h}^{+\infty}\sqrt{\frac{{\mathcal{R}_{GB}^{2}}}{N(r)}}dr \label{11}
\end{eqnarray}
Where the integration in the right side of Eqs.~(\ref{11}) is a constant for a fixed $Q$. Therefore we analytically derive a relation of $\gamma$. 

For $\sqrt{2}>Q>\tilde{Q}$ and $\gamma>0$, we find
\begin{eqnarray}
	\sqrt{\gamma_{0(n+1)}}-\sqrt{\gamma_{0(n)}}=\pi/\int_{r_0}^{+\infty}\sqrt{\frac{{\mathcal{R}_{GB}^{2}}}{N(r)}}dr \label{22}
\end{eqnarray}
For $\sqrt{2}>Q>\tilde{Q}$ and $\gamma<0$, we find
\begin{eqnarray}
	\sqrt{-\gamma_{0(n+1)}}-\sqrt{-\gamma_{0(n)}}=\pi/\int_{r_h}^{r_0}\sqrt{\frac{{\mathcal{R}_{GB}^{2}}}{N(r)}}dr \label{33}
\end{eqnarray}
In this section, we derive an analytical relation of $\gamma_{0(n)}$. We will compare the analytical reslut with numerical data in next section.
\section{ANALYTICAL RESULTS VERSUS NUMERICAL DATA}\label{ssss}
It is interesting to compare the analytical result with numerical data. The numerical solutions can be found by intergrating equation (\ref{scalar eqq}) from horizon to infinity. Using the transformation $\psi\rightarrow\lambda\psi$, we can set $\psi(r_h)=1$ and following the convention in \cite{Brihaye:2019kvj}, we set $r_h=1$. The infinity boundary condition is $\psi(10000)=0$. In Sec.~\ref{section3}, we dervie three different relations (\ref{11}) (\ref{22}) and (\ref{33}) in different situations. We will examine them seperately and suppress subscript "0" in $\gamma_{0(n)}$ for convenience. We use the ratio $R_n\equiv\sqrt{\gamma^{analytical}_n/\gamma^{numerical}_n}$ to exmaine the differecen betwwen the analytical result with numerical data.

For $Q<\tilde{Q}=\sqrt{2(3-\sqrt{6})}\approx1.05$. The coupling parameter can only take positive value. We can use Eqs.~(\ref{1}) to get the analytical result. We consider three cases: $Q=0$, $Q=0.5$ and $Q=1.0$. The results can be seen from Table I

\begin{table}[h]
\centering
\caption{$R_n$ with $Q=0$, $Q=0.5$ and $Q=1.0$ }
\begin{tabular}{|c|c|c|c|c|c|c|c|c|}
\hline
$n$ & 1 & 2 & 3 & 4 & 17 & 18 & 19 & 20 \\
\hline
$R_n(Q=0)$ & 1.1977 & 1.0781 & 1.0483 & 1.0350 & 1.0076 & 1.0071 & 1.0067 & 1.0064 \\
\hline
$R_n(Q=0.5)$ & 1.1553 & 1.0711 & 1.0455 & 1.0334 & 1.0075 & 1.0071 & 1.0067 & 1.0063 \\
\hline
$R_n(Q=1)$ & 1.0261 & 1.0315 & 1.0278 & 1.0240 & 1.0071 & 1.0067 & 1.0064 & 1.0060 \\
\hline
\end{tabular}

\end{table}
From the Table I, we see analytical results agree well with numerical data and the agreement between the analytical result and numerical data improves with increasing of $n$. In fact, the WKB is believed to be valid in the large $n$ limit. However, at this time, the difference between the analytical result and numerical data is already less than $1\%$ when $n=20$.

Let us check the behavior of $\sqrt{\gamma^{numerical}_{n+1}}-\sqrt{\gamma^{numerical}_n}$. In ordr to examine this behavior, we define $\Delta_n\equiv\sqrt{\gamma^{numerical}_{n+1}}-\sqrt{\gamma^{numerical}_n}$. The $Q=0$, $Q=0.5$ and $Q=1.0$ cases are shown in Table II

\begin{table}[h]
	\centering
	\caption{$\Delta_n$ with $Q=0$, $Q=0.5$ and $Q=1.0$}
	\begin{tabular}{|c|c|c|c|c|c|c|c|c|}
		\hline
		$n$ & 1 & 2 & 3 & 4 & 16 & 17 & 18 & 19\\
		\hline
		$\Delta_n(Q=0)$ & 0.6781 & 0.6802 & 0.6803 & 0.6803 & 0.6802 & 0.6802 & 0.6802 & 0.6802\\
		\hline
		$\Delta_n(Q=0.5)$  & 0.6902 & 0.6985 & 0.6999 & 0.7004 & 0.7009 & 0.7010 & 0.7010 & 0.7010\\
		\hline
		$\Delta_n(Q=1)$    & 0.8914 & 0.9039 & 0.9107 & 0.9152 & 0.9227 & 0.9228 & 0.9228 & 0.9229\\
		\hline
	\end{tabular}
\end{table}
The corresponding analytical results can be obtained from Eqs.~(\ref{11}). For $Q=0$, $Q=0.5$ and $Q=1.0$, the corresponding analytical result are $0.6802$ $0.7010$ and $0.9232$. Comparing them with the numerical data in table II, we find analytical result coincide well with numerical data.

Now, we consider the cases where $Q>\tilde{Q}=\sqrt{2(3-\sqrt{6})}\approx1.05$. At this time, the coupling parameter $\gamma$ can take positive or negative value. For the positive branch, we use $R^+_n\equiv\sqrt{\gamma^{analytically}_n/\gamma^{numerical}_n}$  to compare analytically result and numerical data. For the negative branch, we use $R^-_n\equiv\sqrt{\gamma^{analytical}_n/\gamma^{numerical}_n}$. We consider three cases: $Q=1.1$,$Q=1.2$ and $Q=1.4$. The results can be seen from Table III.
\begin{table}[h]
\centering
\caption{$R^+_n$,$R^-_n$ with $Q=1.1$, $Q=1.2$ and $Q=1.4$}
\begin{tabular}{|c|c|c|c|c|c|c|c|c|}
	\hline
	$n$ & 1 & 2 & 3 & 4 & 17 & 18 & 19 & 20 \\
	\hline
	$R^+(Q=1.1)$ & 1.1805 & 1.0846 & 1.0508 & 1.0359 & 1.0076 & 1.0071 & 1.0068 & 1.0064 \\
	\hline
	$R^+(Q=1.2)$ & 1.1598 & 1.0709 & 1.0454 & 1.0333 & 1.0075 & 1.0070 & 1.0067 & 1.0063 \\
	\hline
	$R^+(Q=1.4)$ & 1.1469 & 1.0694 & 1.0448 & 1.0330 & 1.0074 & 1.0070 & 1.0067 & 1.0063 \\
	\hline
    $R^-(Q=1.1)$ & 0.9794 & 0.9976 & 0.9991 & 0.9996 & 1.0000 & 1.0000 & 1.0000 & 1.0000 \\
	\hline
	$R^-(Q=1.2)$ & 0.9204 & 0.9903 & 0.9965 & 0.9982 & 0.9999 & 0.9999 & 0.9999 & 0.9999 \\
	\hline
	$R^-(Q=1.4)$ & 0.5693 & 0.8958 & 0.9602 & 0.9991 & 0.9991 & 0.9992 & 0.9993 & 0.9993 \\
	\hline
\end{tabular}
\end{table}
It is easy to see the difference between analytical results and the numerical data becomes smaller with the increasing of $n$. The difference is already smaller than $1\%$ at $n=20$. These results confirm the validity of the analytic method.

Now, we examine the asymptotic behavior of $\gamma$. For positive branch, we us $\Delta^+_n\equiv\sqrt{\gamma^{numerical}_{n+1}}-\sqrt{\gamma^{numerical}_n}$. For negative branch, we use $\Delta^-_n\equiv\sqrt{-\gamma^{numerical}_{n+1}}-\sqrt{-\gamma^{numerical}_n}$ to investigate the asymptotic behavior of $\gamma$. The results are shown in Table IV 

\begin{table}[h]
\centering
\caption{$\Delta^+_n$,$\delta^-_n$ with $Q=1.1$, $Q=1.2$ and $Q=1.4$}
\begin{tabular}{|c|c|c|c|c|c|c|c|c|}
	\hline
	$n$ & 1 & 2 & 3 & 4 & 16 & 17 & 18 & 19\\
	\hline
	$\Delta^+_n(Q=1.1)$ & 1.3103 & 1.3445 & 1.3433 & 1.3416 & 1.3397 & 1.3397 & 1.3397 & 1.3397\\
	\hline
	$\Delta^+_n(Q=1.2)$ & 1.6587 & 1.6740 & 1.6772 & 1.6783 & 1.6797 & 1.6797 & 1.6797 & 1.6797 \\
	\hline
	$\Delta^+_n(Q=1.1.4)$ & 2.1006 & 2.1286 & 2.1338 & 2.1356 & 2.1378 & 2.1378 & 2.1379 & 2.1379\\
	\hline
	$\Delta^-_n(Q=1.1)$ & 5.1403 & 5.1687 & 5.1729 & 5.1743 & 5.1760 & 5.1760 & 5.1761 & 5.1761 \\
	\hline
	$\Delta^-_n(Q=1.4)$ & 1.5710 & 1.6079 & 1.6132 & 1.6150 & 1.6171 & 1.6171 & 1.6171 & 1.6171 \\
	\hline
	$\Delta^-_n(Q=1.4)$ & 0.2604 & 0.3038 & 0.3166 & 0.3215 & 0.3267 & 0.3268 & 0.3268 & 0.3269 \\
	\hline
\end{tabular}
\end{table}
For positive branch, the corresponding analytical result which can be obtained separately from Eqs.~(\ref{22}) are $1.3396$ $1.6798$ and $2.1380$. For negative branch, the corresponding analytical results which can be obtained separately from Eqs.~(\ref{33}) are $5.1761$ $1.6172$ and $0.3270$. It is easy to see the analytical results agree well with the numerical data.

\section{Conclusion}\label{seciton5}
Recently, by considering the non-minimally coupling of scalar field and Gauss-Bonnet invariant \cite{Doneva:2017bvd,Silva:2017uqg}, they show the black hole can support bound-state hairy scalar configurations. Using numerical method, they show the black hole with scalar hair is characterized by a discrete set $\bar{\eta}\in{[\bar{\eta}^-_n,\bar{\eta}^+_n]}^{n=\infty}_{n=0}$, where $\bar{\eta}$ is the dimensionless coupling parameter between scalar field and Gauss-Bonnet invariant. In the interesting paper \cite{Hod:2019pmb}, they find the $\bar{\eta}^+_n$ which correspond the hairy black-hole-linearized-scalar-field configurations has an asymptotic universal behavior $\sqrt{\bar{\eta}^+_{n+1}}-\sqrt{\bar{\eta}^+_n}$. By using an analytical methods, they provide a compact explanation for this asymptotic universal behavior and the analytical results agree well with the numerical observed.

Motivated by this interesting asymptotic behavior, in this paper, we study the RN black hole with the non-minimally coupling of scalar field and Gauss-Bonnet invariant. We investigate whether the coupling parameter has a similiar asymptotic behavior in this situation. What's more, it is also interesting to examine the validity of the analytical method in this situation. Different from Schwarzschild black hole where the coupling parameter can only take positive value, in this case, the coupling parameter $\gamma$ can take positive or negativie value. By using analytical method, in Sec.~\ref{section3}, we dervie three different relations (\ref{11}) (\ref{22}) and (\ref{33}) in different situations. In Sec.~\ref{ssss}, we compare alytically results with the numerical data (see the data in Table II and IV). We find the analytical results agree well with the numerical data. Therefore, we conclude that the coupling parameter in RN black hole has a similar asymptotic behavior and analytical method give an excellent explanation for this asymptotic universal behavior.

\begin{acknowledgments}
This research was supported by NSFC Grants No. 11775022 and 11873044.
\end{acknowledgments}

\end{document}